\documentclass[review,11pt]{elsarticle} 
\usepackage{algorithm}
\usepackage{algpseudocode}
\usepackage{graphicx}
\usepackage{comment}
\usepackage{amsmath}
\usepackage{subfigure}

\usepackage{booktabs} 
\usepackage{color}
\usepackage[english]{babel}
\usepackage{picinpar}
\usepackage{graphicx}
\usepackage{moresize}
\usepackage{amssymb}
\usepackage{url}
\usepackage{subfigure}
\usepackage{amsmath}

\usepackage{algorithm}

\begin{document}
\begin{frontmatter}

\title{Utilizing Remote Sensing to Analyze Land Usage and Rice Planting Patterns}
\author{Nicholas Milikich}
\address{Department of Computer Science and Engineering\\
University of Notre Dame\\
Notre Dame, IN 46556}

\end{frontmatter}
	\section{Context} \label{sec:context}
\begin{figure}[!htb]
    \centering
   \vspace{-0.05in}
    \includegraphics[width=6cm]{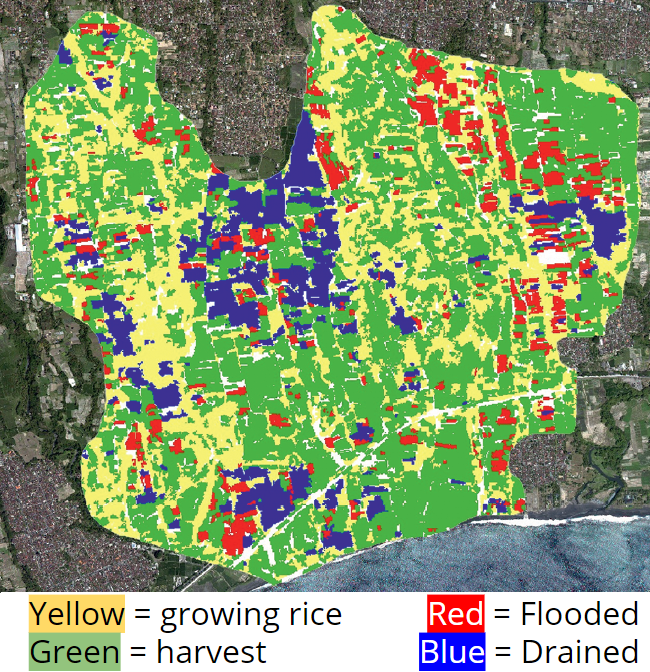}
    \vspace{-0.05in}
    \caption{Rice patches in Bali colored according to planting cycle ~\cite{lansing2017adaptive}}
    \vspace{-0.1in}
    \label{fig:bali}
\end{figure}

The cooperative management of rice terraces in Bali reveal an interesting phenomenon that stems from the feedback loop between human decisions and the ecosystem process~\cite{lansing2017adaptive}.  In particular, a spatial patterning is observed which is heavily reliant on the farmer's decision to plant crops as well as the response from physical environment like pest damage and water shortage. In their paper, Lansing \textit{et al.}~\cite{lansing2017adaptive} proposed an evolutionary game theoretic model to infer particular power laws governing this spatial patterning along the Bali region. Figure \ref{fig:bali} illustrates a snapshot of rice patches in Bali with colors to indicate the different stages of rice growth.

The hypothesis presented by the authors suggest that the complex dichotomy between the human actions and the ecology reaches an optimal state where the harvests are maximized in a non-cooperative game. By experimentation, the authors articulate that the adaptation in a tightly coupled human-natural system can trigger a self-organization pattern~\cite{hasan2016development}. While their developed model is seemingly accurate in its prediction to relate the decision of the Balinese farmers to the consequences of irrigation flows and rice growth, it is not entirely conclusive. The claim that the self-organizing patterns are universal across societies is yet to be explored thoroughly. This opens the door to a new realm of research to expand upon the model further using alternate techniques and apply it at a global scale. 

With the advent of high-precision optical and image processing technologies, satellite-based remote sensing has become a powerful sensing paradigm that can obtain abundant visual features of the objects residing on the earth's surface~\cite{zhang2020transres,zhang2020pqa}. Due to its non-intrusive nature (i.e., not requiring any physical contact), remote sensing is increasingly exploited in scenarios where the detailed analysis of an area cannot be simply performed by modeling or field observations~\cite{venevsky2017emergence}. In our paper, we pose a challenging research question: \textit{Could we leverage remote sensing to extend the analysis proposed by Lansing \textit{et al.} and observe if the organizing patterns are universal across societies?} Thus, we propose an approach to incorporate radar and optical remote sensing to detect and monitor the different stages of the rice cropping cycle with the intention to demystify patches of coordinated planting. The eventual goal is to develop a potent tool to explore the evolution and outcome of ancient cultural traditions and institutions in the management of shared resources.

	\section{Challenges}\label{sec:chhallenge}
Before developing a concrete remote sensing-based platform to study the aforementioned self-organizing patterns, a few potential challenges need to be addressed first~\cite{rashid2020compdrone}. The first major challenge is dealing with cloud covers associated with regions under study. With rice being preferentially planted  in small terraces in tropical humid regions such as Bali, clouds are prominent. Cloud covers are a major obstacle to optical remote sensing and prevent clear analysis of a region~\cite{asner2001cloud}. The second challenge is dealing with low spatial resolutions from traditional satellite imagery. 
Many remote sensing applications need to utilize the historical satellite imagery data to study the spatial and temporal dynamics of an area or phenomenon (e.g., the assessment of land cover changes over time~\cite{al2011assessment}, the study of population migration due to geological changes~\cite{bartholy2004aspects}). Such applications often require the access to a long duration of imagery data (e.g., more than 10 years). Unfortunately, the historic satellite images are often only available in relatively low-resolutions, e.g., the satellite images before 2010 are primarily collected by Landsat 1-7 satellites that only provide low-resolution images (e.g., 30m$\times$30m)~\cite{markham2004landsat}. The third challenge is the sparse data availability from the satellites. It is not possible to have 24/7 high-resolution image coverage of all objects on earth given the current satellite image updating frequency (i.e., from daily to yearly) and communication bandwidth constraints~\cite{yifang2015global}. 
	\newtheorem{myDef}{DEFINITION}
\section{Proposed Methodology} \label{sec:prop}
With an aim to address the above challenges, we intend to develop a robust remote-sensing driven framework able to reliably analyze land usage and rice planting patterns leveraging mathematical analysis and machine learning. As steps of our research in developing the framework, we present a three-fold approach: i) distinguish rice from other land features and crops using data from Sentinel-1, Landsat-7, and Landsat-8; ii) distinguish different stages of the planting cycle using Sentinel-2 (EVI) and Landsat 7; and iii) predict movement of rodents using Landsat-8 (NDVI). After selecting the best tool for each of the three approaches, we intend to meld them together and construct our framework. 

In this particular paper, we discuss a portion of our second approach: distinguish different stages of the rice planting cycle using Sentinel-2.

\subsection{Classifying different stages of the planting cycle using Sentinel-2 (EVI)}
\subsubsection{Applying vector overlay and masking clouds}
\begin{figure}[!htb]
    \centering
   \vspace{-0.05in}
    \includegraphics[width=8cm]{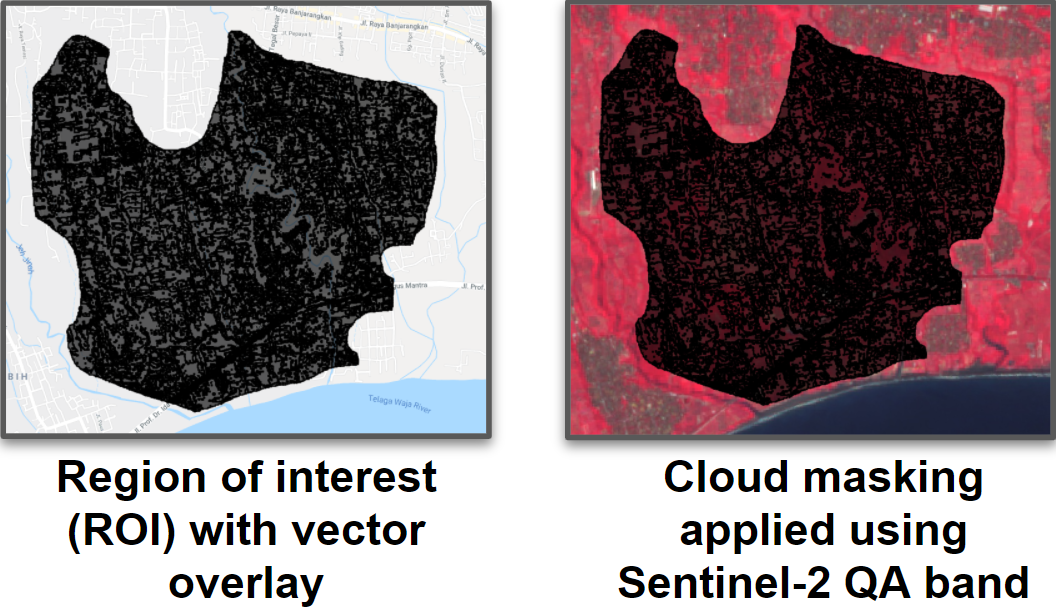}
    \vspace{-0.05in}
    \caption{(left) Region of interest (ROI) with vector overlay; (right) Cloud masking applied using Sentinel-2 QA band}
    \vspace{-0.1in}
    \label{fig:gianyar}
\end{figure}

The first step of our analysis is to select our region of interest (ROI). Since Bali encompasses a substantial area, it is difficult to perform analysis at once using analytics engines like Google Earth Engine (GEE). So first we selected a subarea called Gianyar and proceeded to apply a vector overlay (created in QGIS) in GEE to mark the region as illustrated in Figure~\ref{fig:gianyar}. The second and most important step is to mask the clouds. Conveniently, Sentinel-2A has a special band called the QA60 band that helps to mark clouds and remove them~\cite{tian2018mapping}. The QA60 bitmask has a value of 1024 for opaque clouds and a value of 2048 for cirrus clouds. The application of cloud masking is illustrated in Figure~\ref{fig:gianyar}.

\subsubsection{Computing EVI}
The next step in our analysis is to compute the Enhanced Vegetation Index (EVI) for the region. We utilized information from the NIR (Band 8), RED (Band 4), and BLUE (Band 2) bands to compute the EVI from the Sentinel-2 data using the following formula~\cite{matsushita2007sensitivity}:
\begin{equation}
EVI=2.5\times\frac{NIR-RED}{(NIR+6\times RED-7.5\times BLUE)+1}
\label{evicompute}
\end{equation}
Afterwards, we generated a color palette and mapped the mean EVI for the whole year (this step was just a dry run to test if the prior steps were carried out correctly; the actual analyses are discussed in the subsequent sections).

\subsubsection{Analyzing EVI across the temporal domain}
\begin{figure}[htb]
    \centering
   \vspace{-0.05in}
    \includegraphics[width=12.5cm]{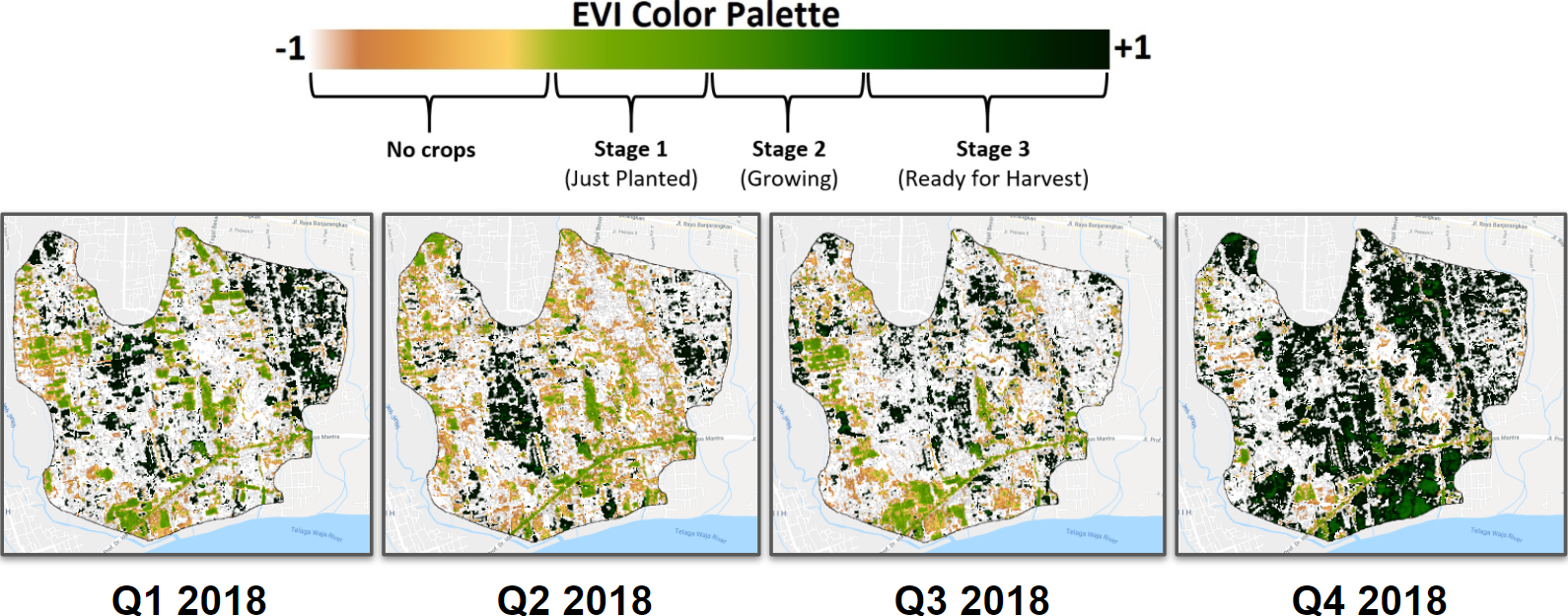}
    \vspace{-0.05in}
    \caption{Variation of Different Rice Stages Along Quarters}
    \vspace{-0.1in}
    \label{fig:temporal}
\end{figure}

To distinguish the different stages of rice production, we carried out EVI analysis across the temporal domain. We kept Gianyar as our ROI and selected four quarters of 2018 as the intervals. We applied unsupervised learning by clustering the computed EVI using Weka \textit{k}-means clustering (setting k=4). After clustering, we correlated our EVI ``clusters" with reference values from~\cite{tian2018mapping,matsushita2007sensitivity} and identified 3 distinct stages of rice (i.e., \textit{just planted}, \textit{growing}, and \textit{ready for harvest}) along with a ``\textit{no crops}" stage (i.e., when the fields are completely barren or doesn't have crops). Figure~\ref{fig:temporal} shows the identified rice stages across the four quarters. 

\subsubsection{Analyzing EVI across the spatial domain}
\begin{figure}[htb]
    \centering
   \vspace{-0.05in}
    \includegraphics[width=12.5cm]{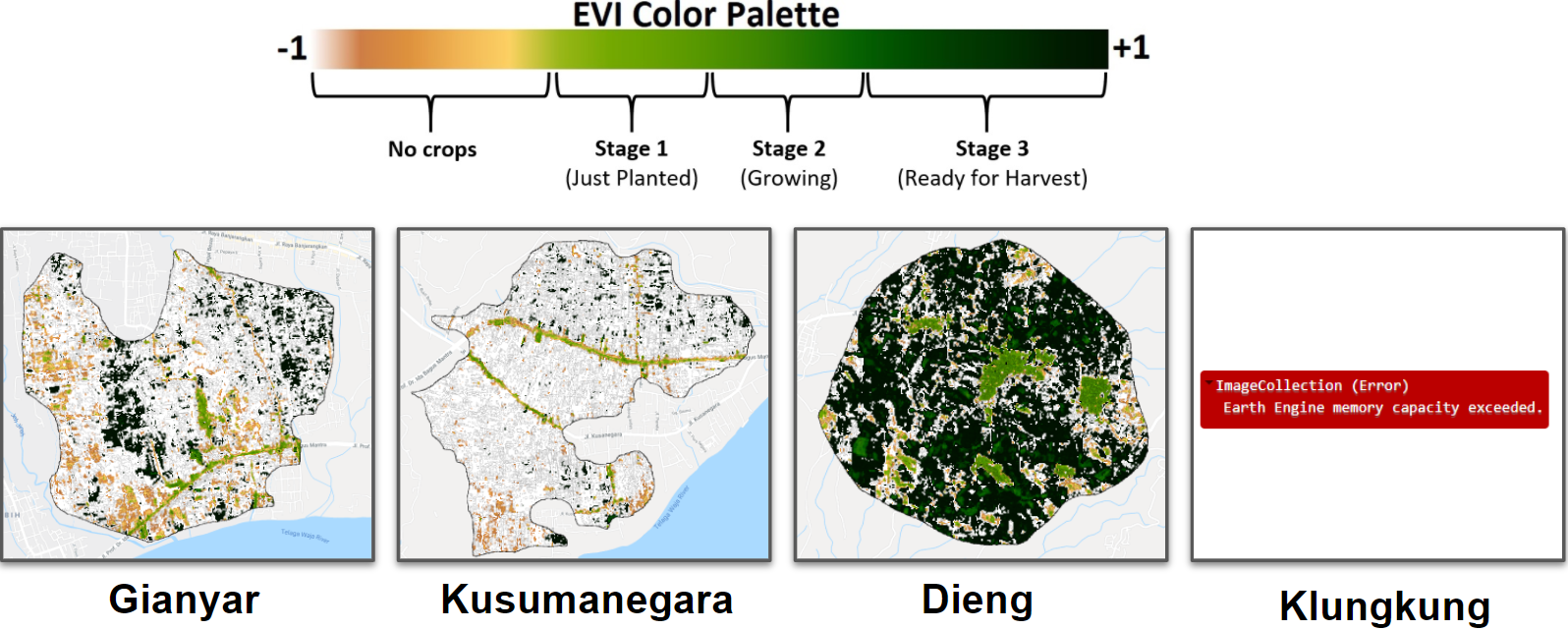}
    \vspace{-0.05in}
    \caption{Recurring Trend of Different Rice Stages Along Different Regions}
    \vspace{-0.1in}
    \label{fig:spatial}
\end{figure}

Next, we applied EVI analysis across different regions and carried out the same steps to cluster and correlate the EVI data. The goal here is to identify the recurring trends of rice production along the different regions. As illustrated in Figure~\ref{fig:spatial}, for example, Dieng has crops ready for harvest for most part of the year, Kusumanegara is mostly barren, while Gianyar has a mixture of all the four stages.

\subsubsection{Synthesizing Results: Comparing with Landsat 7}

	\section{Performance Assessment} \label{sec:perf}
After developing our method, we evaluate the performance of through a cross validation. The evaluation results exhibit acceptable accuracy 
	\section{Discussion and Future Work}
The most notable limitation of this study was the small sample size available. Due to the difficulties associated with data collection, cleaning, and processing, only 51 data points were available for analysis. This severely limited the scope of conclusions that were able to be made: because of the small sample size, it is difficult to validate the model, so it is unclear whether the observed accuracy truly identifies a pattern or is due to chance. In the final predictive model, the coefficients associated with polarity and subjectivity are positive at certain lags and negative at others, suggesting that the model did not capture true patterns but rather noise and chance. It is difficult to manually identify patterns either, as it is impossible to observe long-term trends in public opinion and examine their correlation to the stock market. Compounding this problem is the fact that the time period for which data was collected, March-April 2020, was an extremely volatile time both for the stock market and for public mood. Both signals are likely to have an exceedingly large amount of noise during this time period, and even if trends are able to be identified through this, the period’s uniqueness means it is dubious whether those trends would generalize to “normal,” less volatile time periods.
\bibliographystyle{elsarticle-num}
\bibliography{ref.bib}

\end{document}